\documentclass[seceq]{ptptex}

\usepackage{graphicx,color}
\usepackage{wrapft}

\newcommand{\average}[1]{\left\langle  #1 \right\rangle}

\newcommand{\pq}[1]{\left[{#1}\right]}
\newcommand{\pg}[1]{\left\{{#1}\right\}}

\newcommand{\E}{{\mathrm e}}
\newcommand{\D}{{\mathrm d}}

\title{
Out-of-equilibrium versus dynamical and thermodynamical transitions 
for a model protein
}

\author{
Alberto \textsc{Imparato}$^{1}$, Stefano \textsc{Luccioli}$^{2,3}$, Alessandro \textsc{Torcini}$^{2,3}$
}
\inst{
(1) Dept. of Physics and Astronomy, University of Aarhus,
Ny Munkegade, Building 1520 - DK-8000 Aarhus C, Denmark \\
(2) Istituto dei Sistemi Complessi, CNR, via Madonna del Piano 10, I-50019 Sesto Fiorentino, Italy\\
(3)INFN, Sez. Firenze, and CSDC, via Sansone, 1 - I-50019 Sesto Fiorentino, Italy}

\abst{
Equilibrium and out-of-equilibrium transitions of an off-lattice 
protein model have been identified and studied. In particular, the out-of-equilibrium dynamics of 
the protein undergoing mechanical unfolding is investigated, and by using a work fluctuation relation, the system 
free energy landscape is evaluated. Three different structural transitions are identified along the unfolding pathways. 
Furthermore, the reconstruction of the the free and potential energy profiles in terms of inherent 
structure formalism allows us to put in direct correspondence these transitions with the equilibrium thermal
transitions relevant for protein folding/unfolding. Through the study of the fluctuations of the
protein structure at different temperatures, we identify the dynamical transitions,
related to configurational rearrangements of the protein, which are precursors of the thermal transitions.
}

\PTPindex{02, 07,55}

\begin{document}

\maketitle

\section{Introduction}\label{one}
Biopolymers such as proteins and nucleic acids are paradigmatic examples of complex systems. Similarly to glasses 
and to super-cooled liquids, they are characterized by complex free energy landscapes (FELs), which determine their 
dynamical and thermodynamical properties.  Manipulation experiments on single biomolecules have
made it possible to observe unfolding and refolding trajectories of single proteins \cite{for07}, 
or RNA molecules \cite{rnapap,rnapap1}.  From a theoretical point of view, the unfolding and refolding 
of biomolecules represent  typical stochastic processes where out-of-equilibrium single trajectories of  microscopic systems  
can be observed, since the typical experimental time is much smaller than the typical molecular relaxation time.
When a thermodynamical system is driven far from equilibrium, classical linear response 
theories and other near-equilibrium approximations generally fail. 
However, in recent years some general relations for systems driven far from
equilibrium by large external perturbation have been obtained.  These relations, which are known as fluctuation 
relations \cite{jarz}, go beyond linear response theory valid only in the vicinity of the
equilibrium regime.  On the one hand, single molecule experiments represent excellent test-bed 
for these results obtained in the field  of out-of-equilibrium statistical mechanics \cite{rnapap1}. 
On the other hand, these relations can be used to characterize the thermodynamical properties 
of biomolecules, overcoming the intrinsic out-of-equilibrium nature of the unfolding experiments.
In particular, a fluctuation relation introduced by Hummer and Szabo \cite{HS} can be used to 
estimate the equilibrium free energy landscape of a system as a function of an internal coordinate: 
such a relation has been used to evaluate the FEL of
model \cite{nostro,APZ07,mitter} as well as of real proteins \cite{harris,sbran}.

In two previous papers \cite{nostro}, we have shown that the free energy landscape of 
a model protein can also be evaluated by using the inherent structure (IS) approach, 
a method previously used to characterize the structural-arrest temperature 
in glasses \cite{sastry98} and super-cooled liquids \cite{ang00}. The investigation of 
the IS distributions allows us to give an estimate of the energetic barriers separating the native state from 
the completely stretched configuration along the out-of-equilibrium unfolding trajectories. 
Moreover, the thermal energies to overcome these barriers are related to three temperatures, 
which are quite similar to the temperatures usually employed to 
characterize the thermodynamical transitions associated to protein folding.

Thus, the aim of this paper, is twofold. On the one hand we want to reconstruct via out-of-equilibrium
measurements the free energy landscape of a model protein, by using both a fluctuation relation and IS approach. 
On the other hand, we want to characterize the equilibrium dynamical transitions for the same model protein induced 
by temperature variation, and to compare these with the information gathered via the out-of-equilibrium mechanical manipulations.
 
The paper is organized a follows, in Section  \ref{two} we briefly describe the
model protein used in the present work, and the numerical simulations we perform. In Section \ref{jarz}, we introduce 
the work fluctuation relation and reconstruct the free energy landscape by combining it with out-of-equilibrium 
unfolding simulations. We discuss how the FEL can be evaluated via the IS approach in Section \ref{three}, and make 
a comparison with the results obtained in the previous Section \ref{jarz}.
In Section \ref{four}, we discuss the three characteristic temperatures characterizing 
the thermal unfolding of the protein, and relate them to the structural transition as identified in the 
Sect. \ref{three}. Section \ref{five} is devoted to the analysis of the dynamical transitions
observed for this model protein, which appear to be precursors of the thermal unfolding transitions. 
We conclude and summarize our results in Sect. \ref{six}.

\section{The protein model}\label{two}

The model studied in this paper is a modified version of the 3d off-lattice
model introduced by Honeycutt-Thirumalai \cite{honey} and successively
generalized by Berry {\it et al.} to include a harmonic interaction between
next-neighboring beads instead of rigid bonds \cite{berry}. The model consists 
of a chain of $L$ point-like monomers mimicking the residues of a polypeptidic chain. 
For the sake of simplicity, only three types of residues are considered: hydrophobic (B),
polar (P) and neutral (N) ones.

The intramolecular potential is composed of four terms: a nearest-neighbor harmonic 
potential, $V_1$, intended to maintain the bond
distance almost constant, a three-body interaction $V_2$, which accounts for the
energy associated to bond angles, a four-body interaction $V_3$ corresponding to
the dihedral angle potential, and a long--range Lennard-Jones (LJ) term, $V_4$,
acting on all pairs $i$, $j$ such that $|i-j| >2$, namely
\begin{eqnarray}
V_1 (r_{i,i+1}) &=& \alpha (r_{i,{i+1}}-r_0)^2,
\label{harm}\\
V_2(\theta_i) &=& A \cos(\theta_i) +B \cos(2 \theta_i) - V_0,
\label{bond}\\
V_3(\varphi_i, \theta_i, \theta_{i+1}) &=& C_i [1-S(\theta_i,\theta_{i+1})\cos(\varphi_i))]
 + D_i [1-S(\theta_i,\theta_{i+1})\cos(3 \varphi_i))],
\label{dih}\\
V_4(r_{i,j}) &=& \varepsilon_{i,j} \left( \frac{1}{r_{i,j}^{12}} - \frac{c_{i,j}}{r_{i,j}^6} \right)
\quad .
\label{lj}
\end{eqnarray}
Here, $r_{i,j}$ is the distance between the $i$-th and the $j$-th monomer,
$\theta_i$ and $\varphi_i$ are the bond and dihedral angles at the $i$-th monomer, respectively.
The parameters $\alpha =50$ and $r_0 =1$  fix the strength of the harmonic force and the equilibrium distance between
successive monomers. Both $\alpha$ and $r_0$, as well as all the quantities in the following are expressed in dimensionless units, for
a comparison with physical units see \cite{veit}.

The value of $\alpha$ is chosen to ensure a value for $V_1$
much larger than the other terms of potential in order
to reproduce the stiffness of the
protein backbone. The expression for the bond-angle potential term $V_2 (\theta_i)$ (\ref{bond}) corresponds,
up to the second order, to a harmonic term $\sim k_\theta (\theta_i -\theta_0)^2/2$, where
\begin{equation}
A=- k_{\theta} \frac{cos(\theta_0)}{\sin^2(\theta_0)} ,\qquad
B= \frac{k_{\theta}}{ 4 \sin^2(\theta_0)} ,\qquad
V_0= A \cos(\theta_0) + B \cos( 2\theta_0) \quad ,
\label{param_harm}
\end{equation}
with $k_{\theta} = 20$ and $\theta_0 = 5 \pi/12 \enskip rad$ or $75^o$.

The dihedral angle potential is characterized by three minima for $\varphi = 0$ (associated to
a so-called {\it trans state})  and $\varphi= \pm 2 \pi/3$ (corresponding to {\it gauche states}),
this potential is mainly responsible for the formation of secondary structures. In particular
large values of the parameters $C_i, D_i$ favor the formation of trans state and therefore
of $\beta$-sheets, while when gauche states prevail $\alpha$-helices are formed. The parameters
$(C_i,D_i)$ have been chosen in the following way: if two or more beads among the four defining
$\varphi$ are neutral (N) then $C_i = 0$ and $D_i = 0.2$; in all the other cases
$C_i = D_i = 1.2$. The {\it tapering function} $S(\theta_i,\theta_{i+1})$ has been
introduced in the expression of $V_3$ in order to cure a well known problem in the dihedral
potentials, for more details see~\cite{rampioni,nostro}.
The quantity $S(\theta_i,\theta_{i+1})$ entering in the definition
of $V_3$ has a limited influence on the dynamics apart in proximity of some extreme cases.

The last term $V_4$, introduced to mimic effectively the interactions with the solvent,
is a Lennard-Jones potential, which depends on the type of interacting residues as follows:
if any of the two monomers is neutral the potential is repulsive $c_{N,X} =0$
and its scale of energy is fixed by $\varepsilon_{N,X} = 4$;
for interactions between hydrophobic residues $c_{B,B} =1$ and $\varepsilon_{B,B} = 4$;
for any polar-polar or polar-hydrophobic interaction $c_{P,P} \equiv c_{P,B} = -1$
and $\varepsilon_{P,P} \equiv \varepsilon_{P,B} = 8/3$.

Accordingly, the Hamiltonian of the system reads
\begin{eqnarray}
H = K + V = \sum_{i=1}^L \frac{p_{x,i}^2+p_{y,i}^2+p_{z,i}^2}{2} +\sum_{i=1}^{L-1} V_1(r_{i,i+1}) +
\nonumber\\
+\sum_{i=2}^{L-1} V_2(\theta_i)+
\sum_{i=2}^{L-2} V_3(\varphi_i,\theta_i,\theta_{i+1}) +
\sum_{i=1}^{L-3} \sum_{j=i+3}^{L}  V_4(r_{ij})
\label{hamil}
\end{eqnarray}
where all monomers are assumed to have the same
unitary mass, and consequently the momenta can be defined as $(p_{x,i},p_{y,i},p_{z,i})
\equiv ({\dot x}_i,{\dot y}_i,{\dot z}_i)$.

In the present paper we consider the following sequence of 46 monomers:
\begin{displaymath}
B_9N_3(PB)_4  N_3 B_9  N_3 (PB)_5P
\end{displaymath}

This sequence that has been widely analyzed in the past for thermal 
folding~\cite{honey,guo_thir,guo,veit,berry,evans,kim,kim-keyes} as well
as for mechanically induced unfolding and refolding~\cite{cinpull,lacks,nostro}.
The sequence studied exhibits a four stranded $\beta$-barrel
Native Configuration (NC), which is stabilized by the attractive 
hydrophobic interactions among the $B$ residues (see configuration (a) in Fig. \ref{config.gf}).
In particular the first and third $B_9$ strands, forming the core of the NC,
are parallel to each other and anti-parallel to the second and fourth strand,
namely, $(PB)_4$ and $(PB)_5P$. These latter strands are instead exposed towards 
the exterior due to the presence of polar residues.

In the following we will report simulation results associated to two different
kind of simulation protocols:
equilibrium molecular dynamics (MD) canonical simulations at temperature $T$
performed by integrating the corresponding Langevin equation; steered 
out-of-equilibrium MD simulations intended to mimic the mechanical 
pulling at constant velocity of a protein attached to the cantilever of an atomic force microscope, or 
analogously when trapped in optical tweezers. In both cases the initial state
of the system is taken equal to the native configuration (NC), that we assume
to coincide with the minimal energy configuration.

\section{Fluctuation relation and out-of-equilibrium unfolding}\label{jarz}

Given a system with $L$ particles, characterized by the Hamiltonian $H_0(q)$, 
where $q=\pg{\mathbf r_i, \mathbf p_i}$ is a point in the system phase space, 
we are interested in evaluating the constrained free energy landscape
\begin{equation}
\beta  f_J(Q)\equiv -\ln \pq{ \int \D q\;\delta (Q-Q(q)) \E^{-\beta H_0(q)}},
\label{feldef}
\end{equation} 
where $Q$ is some macroscopic observable, function of the microscopic coordinates $q$.
If the system is driven out of equilibrium by an external potential $U_{z(t)}(Q)$, which depends 
explicitly on $Q$ and on the external parameter $z$, whose temporal evolution is dictated
by the protocol $z=z(t)$, then the FEL (\ref{feldef}) can be obtained via the work fluctuation relation \cite{HS}
\begin{equation}
 \average{\delta (Q-Q(q)) \E^{-\beta W}}=\E^{-\beta \pq{f_J(Q)+U_{z(t)}(Q)}}/Z_0 \quad .
\label{extjar}
\end{equation} 
In Eq. (\ref{extjar}) $W$ is the work exerted on the system by the force associated with the potential $U$, 
and the symbol $\average{\cdot}$ refers to an average over all the
possible stochastic trajectories spanning the system phase space, while the parameter $z(t)$ changes over time. $Z_0$ is the partition function associated with the unperturbed Hamiltonian $H_0(q)$.

\begin{figure}[h]
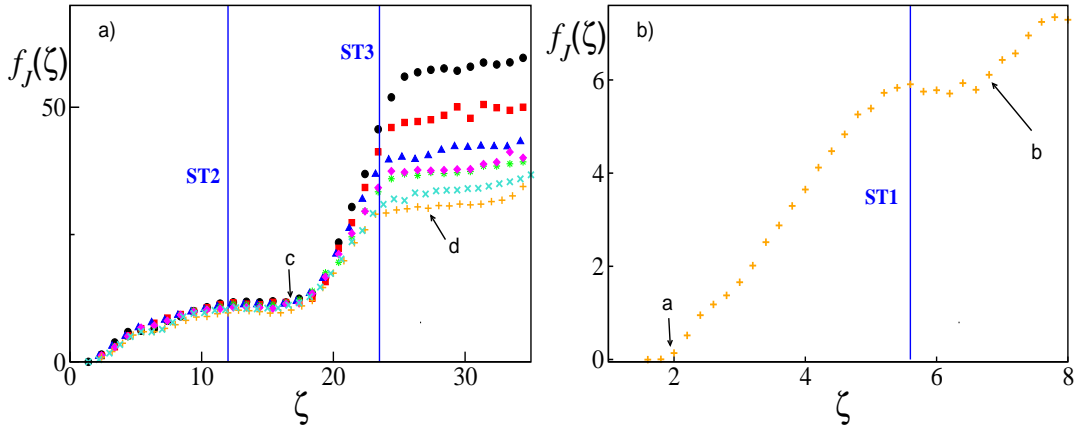

\centerline{
\includegraphics[draft=false,clip=true,height=0.400\textwidth, width=0.50\textwidth]{fig1a.eps}
\includegraphics[draft=false,clip=true,height=0.400\textwidth, width=0.50\textwidth]{fig1b.eps}
}
\caption{Free-energy profiles $f_J$ as a function of the end-to-end distance $\zeta$, 
for the model protein discussed in Sec.~\ref{two}, as obtained by implementing Eq.~(\ref{extjar}), 
with  $T=0.3$, and for different pulling velocities: 
 from top to bottom
$v_p=5 \times 10^{-2}$, $1 \times 10^{-2}$, $5 \times 10^{-3}$,  $5 \times 10^{-4}$,
$2 \times 10^{-4}$, $2 \times 10^{-5}$ and $5 \times 10^{-6}$.
In (b)  an enlargement of the curve for $v_p=5 \times 10^{-6}$ at small $\zeta$ is reported.
The number of different pulling trajectories considered to estimate the profiles ranges between
150 and 250  at the highest velocities
to $28$ at the lowest velocity $v_p=5 \times 10^{-6}$.
The letters (a, b, c, d) indicate the value of $f_J(\zeta)$ corresponding
to typical configurations reported in Fig. \ref{config.gf} 
and the (blue) vertical solid lines the location of the Structural Transitions (STs), see discussion in Sec. \ref{three}.}
\label{figland}
\end{figure}

Typically, the macroscopic observable characterizing the state of a biopolymer under mechanical 
stress is the end-to-end distance $\zeta$ and this is the coordinate we will consider in the following.
Furthermore, we consider a quadratic potential $U_{z(t)}(\zeta)=k/2 (\zeta-z(t))^2$, mimicking the 
effect of the force exerted on the molecule by the cantilever of an atomic force microscope, 
or by optical tweezers, where $z(t)=v_p \times t$, is the equilibrium position of the potential, 
moving with a constant velocity $v_p$.
We perform steered MD unfolding simulations, where one of the free ends of the molecules is kept fixed, 
while the other is pulled by the external force $k(\zeta(t)-z(t))$ associated with the potential $U$. 
Further details on the simulations are given in Ref.~\cite{nostro}. We consider different 
values of $v_p$, and for each of them we simulate a given number of unfolding trajectories. 
For each trajectory, we compute the work $W$ done by the external force on the protein.  Finally, 
in order to estimate $f_J(\zeta)$ from Eq.~(\ref{extjar}) we use the procedure introduced
and discussed in Ref. \cite{IP06}.
The results are plotted in Fig.~\ref{figland} for different pulling velocities $v_p$: we notice that 
as the velocity $v_p$ decreases, the curves collapse onto the same curve which corresponds 
to the best estimate of $f_J(\zeta)$ given by the method described here. In Ref.~\cite{nostro}
we have verified that the lower curve essentially coincides with an independent equilibrium
estimate of the FEL obtained via the weighted histogram analysis method.

\section{Inherent structure formalism and structural transitions}\label{three}

Inherent structures correspond to local minima of the potential energy,
in particular the phase space visited by the protein during its
dynamical evolution can be decomposed into disjoint attraction basins, each corresponding to
a distinct IS. Therefore, the  canonical partition function can be expressed, within the IS formalism, 
as a sum over the non--overlapping basins of attraction, each associated to a specific minimum (IS) $a$
\cite{still2,wales,nakagawa}:
\begin{equation}
Z_{IS}(T) =
\frac{1}{\lambda^{3N^\prime}}
\sum_a {\rm e}^{-\beta V_a} \int_{\Gamma_a}
{\rm e}^{-\beta \Delta V_a (\Gamma)} d \Gamma  =
\sum_a {\rm e}^{-\beta  [V_a+R_a(T)]}
\label{zeta}
\end{equation}
where $N^\prime$ is the number of degrees of freedom of the system,
$\lambda$ is the thermal wavelength, $\Gamma$ represents one of the possible conformations of
the protein within the basin of attraction of $a$,
$V_a$ is the potential energy associated to the minimum $a$, $\Delta V_a (\Gamma)=V(\Gamma)-V_a$ and
$R_a(T)$ the vibrational free energy due to the fluctuations around the minimum.

\begin{figure}[t]
\centerline{
\includegraphics[draft=false,clip=true,height=0.60\textwidth, width=0.80\textwidth]{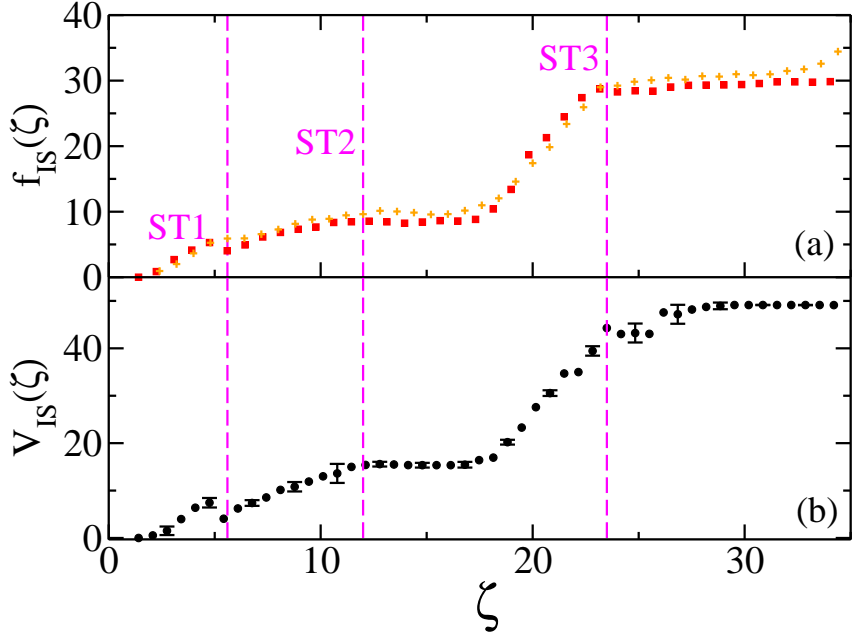}
}
\caption{(Color online) (a) Free energy profile $f_{IS}(\zeta)$ as a function of the end-to-end distance, (b) average
potential energy $V_{IS}(\zeta)$ vs $\zeta$. The dashed vertical lines indicate the location
of the three structural transitions. All the data refer to a data bank of ISs obtained
via out-of-equilibrium steered MD simulations mimicking mechanical protein 
unfolding performed at $T=0.3$ with $v_p=5\times10^{-4}$. In the upper panel (a) it is
reported for comparison also $f_J(\zeta)$ (orange pluses) obtained via the fluctuation relation procedure  
 with $v_p=5\times10^{-6}$.
}
\label{f2}
\end{figure}

The free energy of the whole system at equilibrium is simply given by
$ f_{IS}(T)= - T \ln[Z_{IS}(T)]$. However, in order to construct
a free energy landscape as a function of a parameter characterizing the
different IS, like e.g. the end-to-end
distance $\zeta$, it is necessary to define a partition function
restricted to ISs with an end-to-end distance within a narrow interval
$[\zeta; \zeta +d \zeta]$
\begin{equation}
Z_{IS}(\zeta, T) =
{\sum_a}^\prime {\rm e}^{-\beta [V_a+R_a(T)]}
\label{zetaprime}
\end{equation}
where the $\sum^\prime$ indicates that the sum is not over the whole ensemble
of ISs $\{ a \}$ but restricted.
The free energy profile as a function of $\zeta$ can be obtained by the
usual relationship $f_{IS}(\zeta,T)= - T \ln[Z_{IS}(\zeta, T)]$;
while the average potential energy, corresponding to ISs
characterized by a certain $\zeta$, can be estimated as follows:
\begin{equation}
V_{IS}(\zeta,T) = \frac{{\sum_a}^{\prime} V_a \enskip {\rm e}^{-\beta [V_a+R_a(T)]}}{Z_{IS}(\zeta,T)} \qquad .
\label{VIS_RIS}
\end{equation}

In order to built a data bank containing the different ISs, we have performed mechanical unfolding 
simulations of the protein at different temperatures via steered Langevin MD integration schemes.
The data bank contains $3,000 - 50,000$ ISs depending on the examined temperature as detailed 
in Ref.~\cite{nostro}. It is worth to notice that the ISs have been collected by following
out-of-equilibrium trajectories induced by mechanical manipulation on the NC at a velocity
$v_p=5\times10^{-4}$. As shown in Fig.~\ref{f2} (a), the free energy profile reconstructed
with this approach is almost coincident with the best estimate of $f_{J}(\zeta)$. However, the 
fluctuation relation procedure, described in the previous Section, requires a pulling velocity  which is 
two orders of magnitude smaller in order to obtain a reliable reconstruction.

Moreover, referring to Fig.~\ref{figland} and Fig.~\ref{f2} (a), it is possible to identify the structural 
transitions (STs) induced by the pulling experiment. As shown in Fig.~\ref{figland} (b), the free energy profile exhibits a
clear minimum in correspondence of the end-to-end distance of the
NC (namely, $\zeta_0 \sim 1.9$). In more detail, up to $\zeta \sim 5.6$, the protein remains in
native-like configurations characterized by a $\beta$-barrel made up of 4 strands, while the
escape from the native valley is signaled by the small dip at $\zeta \sim 5.6$ and it is
indicated as ST1 in Fig. ~\ref{figland} (b) and Fig.~\ref{f2} (a).

\begin{figure}[t]
\centerline{
\includegraphics[draft=false,clip=true,height=0.20\textwidth,width=0.50\textwidth]{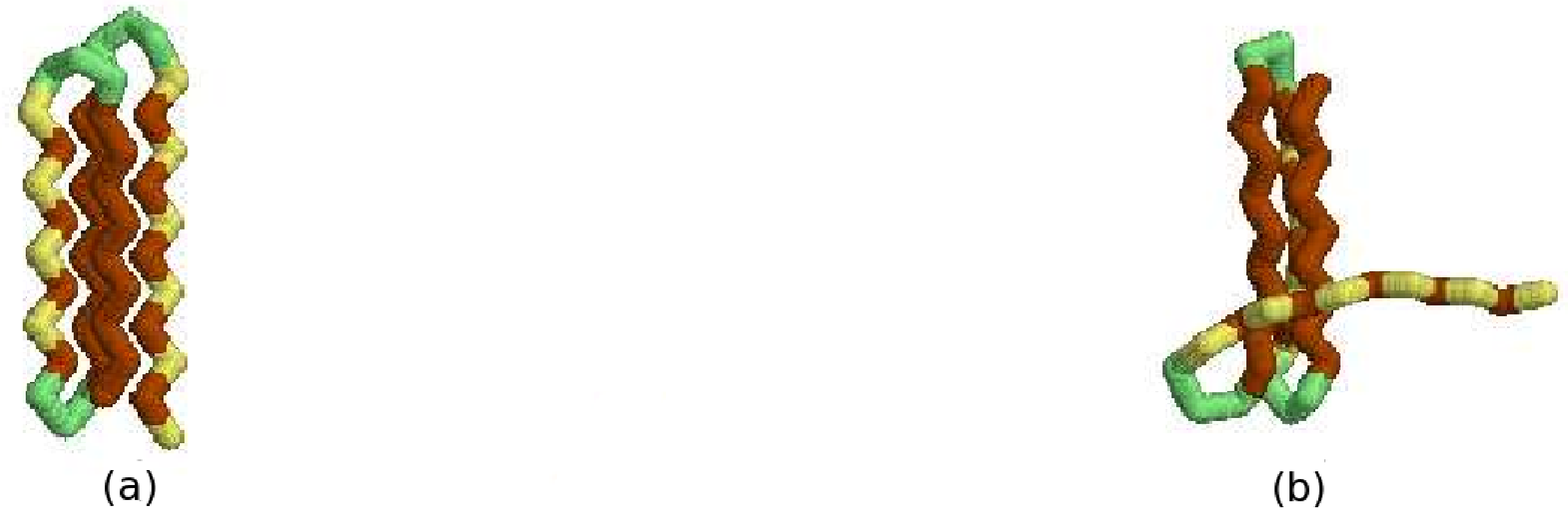}
\includegraphics[draft=false,clip=true,height=0.20\textwidth,width=0.50\textwidth]{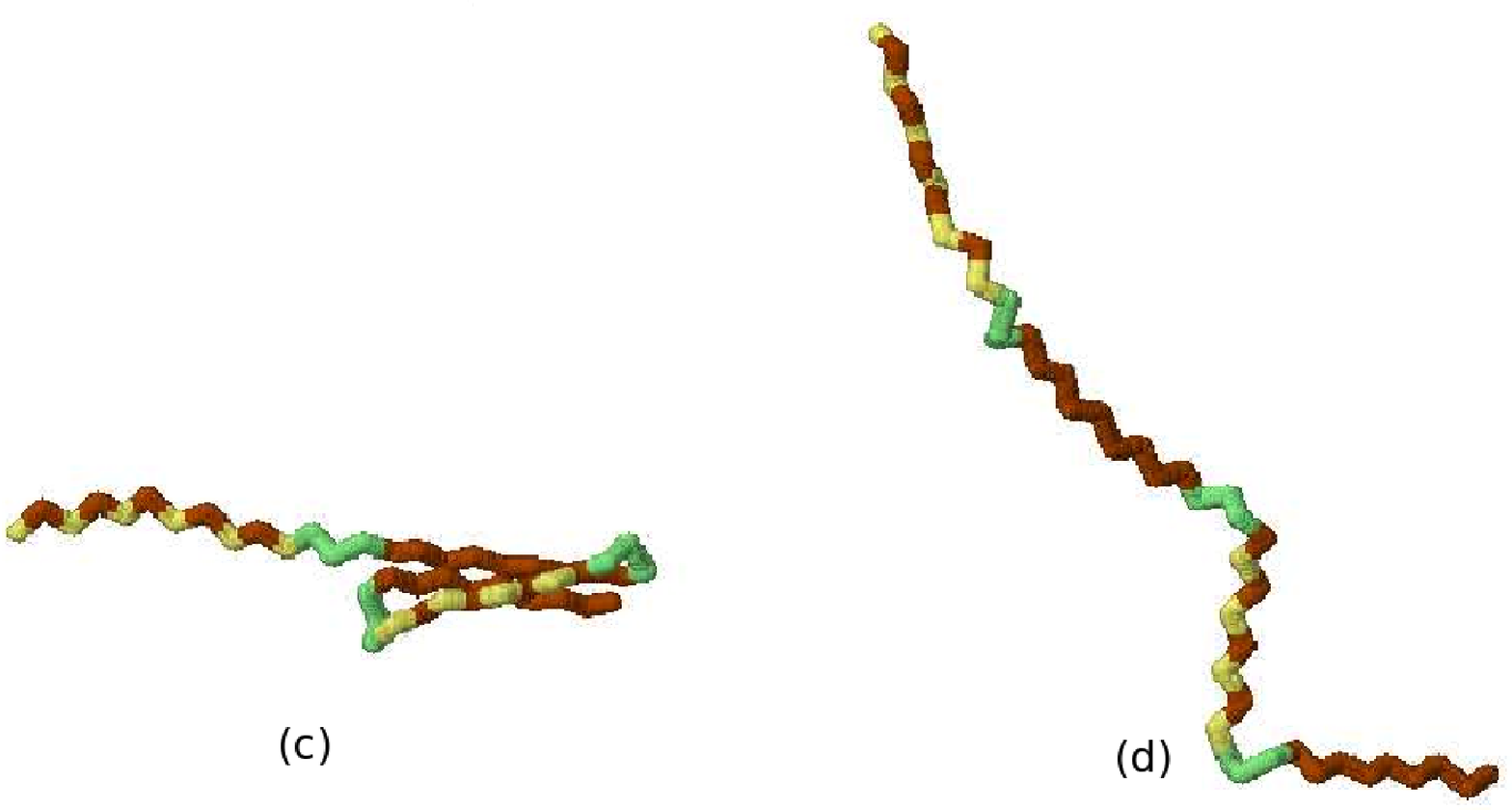}
}
\caption{(Color online) Typical configurations of the model protein along an unfolding trajectory 
driven by the mechanical force discussed in Section \ref{jarz}, with $T=0.3$. The NC (a) has $\zeta_0 \sim 1.9$;
the other configurations are characterized by $\zeta=6.8$ (b), $\zeta=16.8$ (c), and $\zeta=27.1$ (d).
The beads of type $N$, $B$, and $P$ are colored in green,
red and yellow, respectively. 
}
\label{config.gf}
\end{figure}

For $\zeta > 6 $ the configurations are
characterized by an almost intact core (made of 3 strands) plus a stretched
tail corresponding to the pulled fourth strand (see configuration (b)
in Fig. \ref{config.gf}).
The second ST amounts to pull the strand $(PB)_5P$ out of the barrel leading
to configurations similar to (c) reported in Fig. \ref{config.gf}.
In the range $13 < \zeta < 18.5$ the curve $f_{IS}(\zeta)$ appears as essentially flat,
thus indicating that almost no work is needed to completely stretch the tail once detached from
the barrel. The pulling of the third strand (that is part
of the core of the NC) leads to a definitive destabilization of the $\beta$-barrel. This transition is denoted
as ST3 in Fig. \ref{f2} (a). The second plateau in $f_{IS}(\zeta)$ corresponds to protein structures made
up of a single elongated strand (an example of this state is configuration (d) in Fig. \ref{config.gf}).

\section{Thermodynamical transition temperatures}\label{four}

The main thermodynamic features of a protein can be summarized with reference to
three different transition temperatures \cite{wales,evans,kim-keyes,tlp,nostro}:
the collapse temperature $T_\theta$ discriminating between phases
dominated by random-coil configurations rather than collapsed ones;
the folding temperature $T_f$, below
which the protein stays predominantly in the native valley;
and the glassy temperature $T_g$ indicating the freezing of
large conformational rearrangements \cite{nakagawa}.
Following the procedures reported in Ref. \cite{tlp},
we have determined these temperatures and obtained
$T_\theta = 0.65(1)$, $T_f = 0.255(5)$, and $T_g=0.12(2)$. These values
are in good agreement with those reported in \cite{evans,kim-keyes},
where $T_f$ and $T_g$ have been identified via different protocols.

Then, we can try to put in correspondence the three unfolding stages previously
discussed in Sec. 4 with thermodynamical aspects of the protein folding. In particular,
by considering the energy profile $V_{IS}(\zeta)$ reported in Fig.~\ref{f2} (b), an 
energy barrier $\Delta V_{IS}$ and a typical transition temperature $T_t = (2 \Delta V_{IS})/(3N)$,
can be associated to each of the STs. The first transition ST1 corresponds to a barrier
$\Delta V_{IS} = 8(1)$ and therefore to $T_t = 0.11(1)$, that, within error bars, essentially coincide with $T_g$.
For the ST2 transition to occur, the barrier to overcome is $\Delta V_{IS} = 16(1)$ and this is associated
to a temperature $T_t = 0.23(2)$ (slightly smaller than $T_f$). The energetic cost to completely stretch
the protein is $50(1)$   that corresponds to a transition temperature $T_t = 0.72(1)$, that is not too far from
the $\theta$-temperature given above.
At least for this specific sequence, our results indicate that the observed out-of-equilibrium STs
induced by mechanical pulling can be put in direct relationship with equilibrium thermal transitions 
usually characterizing the folding/unfolding processes.

\section{Dynamical transitions and structure fluctuations}\label{five}

In the analysis of equilibrium properties of heteropolymers,
an abrupt deviation of the structural mean square displacement from a
linear temperature dependence is usually associated to a {\it dynamical transition}~\cite{doster}.
Fluctuations of the protein structure at equilibrium have been studied experimentally
via elastic incoherent neutron scattering as well as M\"ossbauer absorption spectroscopy~\cite{bicout,parak,doster}. 
These studies indicate the existence of different dynamical regimes separated
by dynamical transitions~\cite{doster}, in particular for hydrated proteins powders,
a first non-linear enhancement of the mean square displacement with the temperature is observed
around 150 K, and a second one around 240 K. The first dynamical transition is associated
to torsional motions and observable also for dehydrated or solvent-vitrified system, while
the second one is related to the onset of small-scale libration motions of side-chains
induced by water at the protein surface.

Fluctuations of the protein structure at a certain temperature
can be characterized in terms of the following indicator~\cite{nakagawa}
\begin{equation}
\langle \langle \Delta u^2 \rangle \rangle = \frac{1}{L} \sum_{i=1}^L \Delta u^2_i
\qquad {\rm where} \qquad
\Delta u^2_i = \langle d_{i,CM}^2 \rangle - \langle d_{i,CM} \rangle^2 \qquad ,
\label{indicator}
\end{equation}
associated to the fluctuations of
the distance $d_{i,CM}$ between the $i$-th residue and the center of mass of the
protein, the symbols $\langle \cdot \rangle $ refer to temporal averages, while 
$\langle \langle \cdot \rangle \rangle$ to an average over all the beads composing the heteropolymer.

In the present case, we have estimated $\langle \langle  \Delta u^2 \rangle \rangle$ for various temperatures 
$T$ by performing equilibrium unfolding MD canonical simulations and by 
following the protein trajectory for a time $t= 500,000$.
In Ref.~\cite{nakagawa} the authors have shown for off-lattice G${\mathrm{\bar o}}$ models, 
reproducing the $B_1$ domain of protein G, the existence of a dynamical
transition temperature $T_D \sim 0.4 \times T_f$ denoting the onset of large scale fluctuations.
As already mentioned in the previous Section, our model (as real proteins) is characterized by
three different transition temperatures, at variance with the G${\mathrm{\bar o}}$ model examined in 
Ref.~\cite{nakagawa}, where the folding and the collapse temperature coincide.
It is therefore quite instructive to examine how many dynamical transitions
are present in our model and their location in temperature.

\begin{figure}[t]
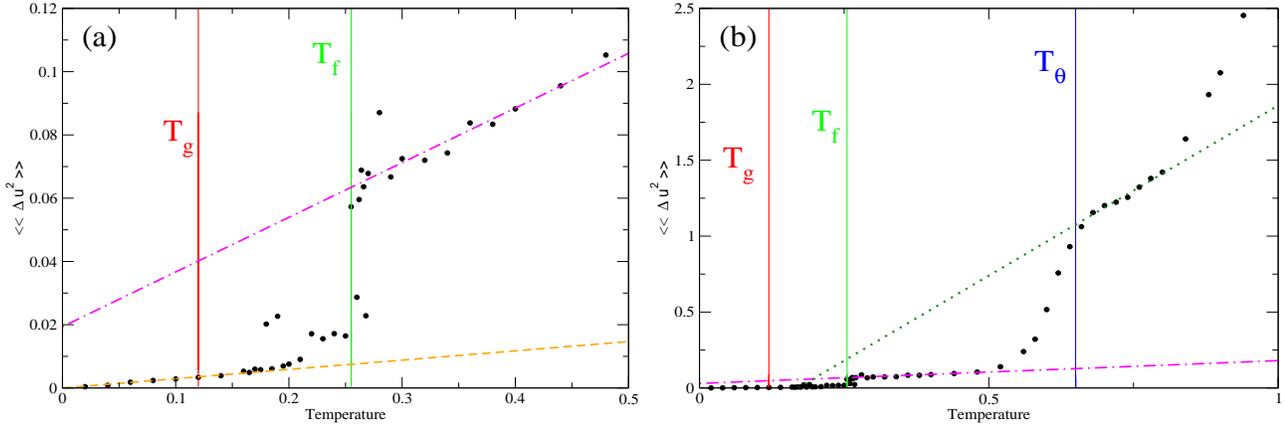

\centerline{
\includegraphics[draft=false,clip=true,height=0.40\textwidth, width=0.60\textwidth]{fig4a.eps}
\includegraphics[draft=false,clip=true,height=0.40\textwidth, width=0.60\textwidth]{fig4b.eps}
}
\caption{(Color online)
Variance $\langle \langle \Delta u^2 \rangle \rangle$ versus temperature $T$, as estimated from equilibrium 
unfolding simulations of duration $t = 500,000$. The (black) dots are the results of the simulations,
while in (a) the dashed (orange) line corresponds to $\langle \langle \Delta u^2 \rangle \rangle = \gamma_1 T$ with $\gamma_1=0.02935$
and the dash-dotted (magenta) line to $\langle \langle \Delta u^2 \rangle \rangle = (\gamma_1 + \gamma_2) T + d_0$ with
$\gamma_2=0.1434$. In (b) the dotted (green) line refers to  
$\langle \langle \Delta u^2 \rangle \rangle = (\gamma_1 + \gamma_2 + \gamma_3) T + d_1$ with $\gamma_3=2.067$.
}
\label{f1}
\end{figure}

From Fig. \ref{f1}~(a) it is clear that $\langle \langle  \Delta u^2 \rangle \rangle$ exhibits a
linear behaviour until $T_{D1} \sim 0.2 \sim 0.78 \times T_f$ where a sharp increase
takes place. At $T < T_{D1}$ the equilibrium dynamics is simply characterized by
small harmonic oscillations of the beads around their equilibrium positions.
Therefore by applying the theorem of equipartition of energy to the corresponding
potential term ($\ref{harm}$) we expect that
\begin{equation}
\langle \langle \Delta u^2 \rangle \rangle = \frac{3}{2 \alpha} \frac{L-1}{L} T = \gamma_1 T \qquad {\rm for }
\enskip T \le  T_{D1} \quad,
\label{lin1}
\end{equation}
as indeed verified (see Fig. \ref{f1}(a)). By approaching the folding temperature
there is a strong nonlinear enhancement of $\langle \langle \Delta u^2 \rangle \rangle$ due to
a configurational rearrangement of the protein structure, which can be associated
to an activation process which leads the protein to
cross the free energy barrier at ST2 (see Fig.~\ref{f2}(a)).

Moreover just above $T_f$ a second linear regime is observable. This is due to angular 
oscillations around their equilibrium positions $\theta_0$, therefore by applying 
equipartition to the terms $V_1$ and $V_2$ in our model we derive the following
dependence
\begin{equation}
\langle \langle \Delta u^2 \rangle \rangle \sim \gamma_1 T + \frac{3}{k_\theta} \frac{L-2}{L} T = (\gamma_1 + \gamma_2) T
\qquad {\rm for } \enskip T_f \le T \le  T_{D2}
\label{lin2}
\end{equation}
and again this linear behaviour is in good agreement with the data up to
$T_{D2} \sim 0.5 \sim 0.77 \times T_\theta$ (see Fig. \ref{f1}(a)). 
In this temperature range, the protein is partially unfolded, it is no more in
native-like configurations, and the degrees of freedom associated to
bending fluctuations, involving three consecutive beads, are now activated.
At the temperature $T_{D2}$ we observe a second dynamical
transition involving large configurational fluctuations.
This dynamical transition can be considered as a precursor of the collapse 
transition, characterized by the complete unfolding of the protein and associated to the crossing of the free energy barrier at ST3 in Fig.~\ref{f2}(a).

As a matter of fact just above $T_\theta$ an {\it almost} linear regime
is observable in a narrow temperature interval, namely $0.68 \le T \le 0.80$,
we believe that this further linear regime is due to fluctuations of the dihedral angles
entering in the potential term $V_3$.  We found that, in this temperature range
the linear increase is characterized by a slope $(\gamma_1 + \gamma_2 + \gamma_3) \sim 2.24$, 
leading to an estimate for the new contribution  $\gamma_3 \sim 2.07$ 
which is of the order of $3 / D_i = 2.5$ (by assuming that no neutral bead is 
involved in the oscillating dihedral angles).  These fluctuations are indeed more 
collective since they involve four consecutive beads. A third dynamical transition
appears to take place around $T_{D3} \sim 0.82$. This latter transition is probably
related to fluctuations involving large part of the protein.

Therefore each of the observed dynamical transitions is 
first characterized by small oscillations around some {\it typical
equilibrium configuration} (this corresponds to the linear regime)
followed by larger fluctuation induced by the breaking of
hydrophobic bonds and leading to a new {\it equilibrium configuration}
of the protein (this phase is characterized by an abrupt increase in the protein fluctuations). 
Once the protein is rearranged one observes another linear regime due to
the activation of a different set of degrees of freedom, which
were previously hindered by the hydrophobic interactions. At temperatures
$ T \le T_{D1}$ the protein stays essentially in 
tightly packed native-like configuration and the only allowed
oscillations are those of the beads around their equilibrium positions.
At $ T_f < T \le T_{D2}$ the protein visits a different sets of
less packed equilibrium configurations of the free energy, which are  still 
characterized by a native core essentially intact. In this regime the bending
oscillations of three consecutive beads become possible and are present
together with harmonic oscillations of each bead. The transition at $T_\theta$ leads essentially to 
configurations  almost completely stretched where fluctuations involving four
consecutive beads (defining a dihedral angle) are now also activated.

From the analysis of $\langle \langle \Delta u^2 \rangle \rangle$ we have no indication of the
glassy transition occurring at $T_g$, apart some fluctuation taking
place just above $T_g$ as shown in Fig.~\ref{f1}(a). This is probably due to the fact that 
 we have traced the dynamics of the protein for too short time windows.
However, by increasing by a factor five the integration time, we do not
observe substantial modifications in the behavior of $\langle \langle \Delta u^2 \rangle \rangle$.

\section{Conclusions}\label{six}

In the present paper, we have discussed how the FEL of a model protein 
driven out of equilibrium can be estimated by exploiting two different methods: 
namely, we applied a work fluctuation relation and the IS approach.
The results obtained with the two methods compare well, although the IS approach 
provides a reliable estimate of the FEL already at larger pulling 
velocities compared to the first method.

The FEL reveals three structural transitions along the unfolding pathways.
By evaluating the potential energy landscape, we are able to assign a characteristic 
temperature to each of these structural transitions. Such temperatures compare well with 
the temperatures characterizing the {\it thermal} (un)folding of the molecule. 
Finally we analyze in detail the equilibrium structure fluctuations which mark the 
folding and collapse thermal transitions. Inspection of these fluctuations' variance 
allows us to identify the dynamical transitions which turn out to be precursors 
of the  two corresponding structural transitions (namely, ST2 and ST3).

In conclusion, our work provides strong evidence that, at least for the present 
protein model, the mechanical out-of-equilibrium unfolding pathways can be reconciled 
with the thermal folding and unfolding ones, provided that one performs a detailed 
analysis of the relevant quantities, namely the free and potential energy landscapes, 
and the thermal transition temperatures.

\section*{Acknowledgements}
We would like to thank S. Lepri for useful suggestions and discussions.
This work has been partially supported by the Italian project ``Dinamiche cooperative in
strutture  quasi uni-dimensionali'' N. 827 within the CNR programme ``Ricerca
spontanea a tema libero''.
AI gratefully acknowledges support for computing resources from Danish
Centre for Scientific Computing (DCSC).

%


\begin{thebibliography}{99}

\bibitem{for07} J.R. Forman and J. Clarke, Curr. Opin. Struct. Biol. {\bf 17} (2007), 58.

\bibitem{rnapap} B. Onoa B {\it et al},  Science {\bf 299} (2003) 1892.

\bibitem{rnapap1} D. Collin {\it et al}, Nature {\bf 437} (2005) 231.

\bibitem{jarz} C. Jarzynski, Phys. Rev. Lett. {\bf 78} (1997) 2690; 
C. Jarzynski, Phys. Rev. E {\bf 56} (1997) 5018; 
G.E. Crooks, J. Stat. Phys. {\bf 90} (1998) 1481; G.E. Crooks, Phys. Rev. E {\bf 60} (1999) 2721.

\bibitem{HS} G. Hummer and A. Szabo, Proc. Natl. Acad. Sci. USA. { \bf 98} (2001) 3658.

\bibitem{APZ07}
A. Imparato, A. Pelizzola, and M. Zamparo, Phys. Rev. Lett. {\bf 98} (2007) 148102; 
A. Imparato, A. Pelizzola, M. Zamparo, J. Chem. Phys {\bf 127} (2007) 145105.

\bibitem{nostro} A. Imparato, S. Luccioli, and A. Torcini, \PRL{99,2007,168101};
S. Luccioli, A. Imparato, and A. Torcini, \PRE{78,2008,031907}.

\bibitem{mitter} S. Mitternacht, S. Luccioli, A. Torcini, A. Imparato, A. Irb\"ack, Biophys. J. {\bf 96} (2009) 429.

\bibitem{harris} N. C. Harris, Y. Song, and C.-H. Kiang, Phys. Rev. Lett. {\bf 99} (2007) 068101.

\bibitem{sbran}  A. Imparato, F. Sbrana, and M. Vassalli, Europhys. Lett. {\bf 82} (2008) 58006.

\bibitem{sastry98} S. Sastry, P. G. Debenedetti, and F. H. Stillinger, Nature (London) {\bf 393} (1998) 554.

\bibitem{ang00} L. Angelani, R. Di Leonardo, G. Ruocco, A. Scala, and F. Sciortino, Phys. Rev. Lett. {\bf 85} (2000) 5356.

\bibitem{honey} J.D. Honeycutt and D. Thirumalai, {Proc. Natl. Acad. Sci. U.S.A.} {\bf 87} (1990) 3526.

\bibitem{berry} R.S. Berry, N. Elmaci, J.P. Rose, and B. Vekhter, {Proc. Natl. Acad. Sci. U.S.A.} {\bf 94} (1997) 9520.

\bibitem{veit} T. Veitshans, D. Klimov, and D. Thirumalai, {Folding \& Design} {\bf 2} (1997) 1.

\bibitem{rampioni} A. Rampioni, {\it Caratterizzazione del panorama energetico
di piccoli peptidi al variare della loro lunghezza}, PhD Thesis (Firenze, 2005)

\bibitem{kim-keyes} J. Kim and T. Keyes, {J. Phys. Chem. B} {\bf 111} (2007) 2647

\bibitem{guo_thir} Z. Guo and D. Thirumalai, Biopolymers, {\bf 36} (1995) 83.

\bibitem{guo} Z. Guo and C.L. Brooks III, Biopolymers, {\bf 42} (1997) 745-757.

\bibitem{evans} D.A. Evans and D.J. Wales, {J. Chem. Phys} {\bf 118} (2003) 3891.

\bibitem{kim} J. Kim, J.E. Straub, and T. Keyes, \PRL{97,2006,050601}.

\bibitem{lacks} D.J. Lacks, Biophys. J. {\bf 88} (2005) 3494.

\bibitem{cinpull} F.-Y. Li, J.-M. Yuan, and C.-Y. Mou, \PRE{63,2001,021905}.

\bibitem{IP06}  A. Imparato, L. Peliti, J. Stat. Mech. (2006) P03005.

\bibitem{wales} D.J. Wales, {\em Energy Landscapes}, Cambridge University Press, Cambridge, 2003.

\bibitem{still2} F.H. Stillinger and T.A. Weber,
{\em Science} {\bf 225} (1984) 983.

\bibitem{nakagawa} N. Nakagawa and M. Peyrard, {Proc. Natl. Acad. Sci. USA}
{\bf 103} (2006) 5279; \PRE{74,2006,041916}.

\bibitem{tlp} A. Torcini {\it et al.} {J. Biol. Phys.} {\bf 27} (2001) 181; 
L. Bongini {\it et al.} \PRE{68,2003,061111}.

\bibitem{wales2004} D.A. Evans and D.J. Wales, {J. Chem. Phys} {\bf 121} (2004) 1080.

\bibitem{doster} W. Doster, \JL{Eur. Biophys. J.,37,2008,591} 

\bibitem{bicout} D.J. Bicout and G. Zaccai,  \JL{Biophys. J.,80,2001,1115} 

\bibitem{parak} F.G. Parak,  \JL{Curr. Opin. Struct. Biol.,13,2003,552} 




 
\end{thebibliography}
\end{document}